\documentclass[graybox]{svmult}

\usepackage{mathptmx}       
\usepackage{helvet}         
\usepackage{courier}        
\usepackage{type1cm}        
\usepackage{graphicx}        
\usepackage{multicol}        
\usepackage[bottom]{footmisc}

\usepackage{amsmath}
\usepackage{amssymb}

\usepackage{boxedminipage}
\usepackage{psfrag}

\usepackage{url}

\newcommand{\subsubsubsection}{\heading}
\newcommand{\heading}[1]{\medskip\noindent{\bf #1}} 

\renewcommand{\index}{}

\newcommand{\pn}{\beta}

\newcommand{\remove}[1]{}
\newcommand{\opt}{\mathrm{OPT}}
\newcommand{\argmax}{\arg \max}

\newcommand{\bidders}{{\cal B}}
\newcommand{\aecpm}{a}
\newcommand{\eff}{e}
\newcommand{\ignore}[1]{}

\newcommand{\reals}{{\mathbb R}}
\newcommand{\half}{{\frac 1 2}}
\newcommand{\set}[1]{ \{ {#1} \} }
\newcommand{\eps}{\epsilon}

\newcommand{\ctr}{\alpha} 
\newcommand{\userctr}{\ctr}

\newcommand{\cl}{{\rm clicks}}
\newcommand{\tcl}{{\rm traffic}}

\newcommand{\cost}{{\rm cost}}
\newcommand{\tcost}{{\rm spend}}

\newcommand{\bid}{{\rm b}}
\newcommand{\G}{G} 
\newcommand{\K}{K} 
\newcommand{\Q}{Q} 
\newcommand{\edges}{E} 
\newcommand{\poss}{p} 
\newcommand{\bctr}{\pn} 
\newcommand{\B}{U} 
\newcommand{\bidd}{{\cal B}} 
\newcommand{\bidv}{{\bf a}} 
\newcommand{\bidvd}{{\cal A}} 
\newcommand{\iposs}{{\rm pos}}
\newcommand{\OPT}{\Omega} 
\newcommand{\lan}{V} 
\newcommand{\bigbidset}{{\cal I}} 
\newcommand{\bidset}{I} 
\newcommand{\agland}{{\cal V}}
\newcommand{\tpoints}{N}
\newcommand{\kbidding}{{\sc Budget Optimization} } 

\newcommand{\m}{m}

\newcommand{\DD}{D}

\newcommand{\carr}{\mathbf{c}}

\newcommand{\bidgt}{\succ}

\newenvironment{mechanism}[1]{
\begin{center}
\begin{svgraybox}
\begin{tabular}{p{4.2in}}
{\bf #1}\\
}{
\end{tabular}
\end{svgraybox}
\end{center}
}

\begin{document}

\title*{Algorithmic Methods for Sponsored Search Advertising} 

\author{Jon Feldman and S. Muthukrishnan}
\institute{Jon Feldman \at Google, Inc., 76 $9^{\mathrm{th}}$ Avenue, $4^{\mathrm{th}}$ Floor, New York, NY, 10011. \email{jonfeld@google.com}
\and  S. Muthukrishnan \at Google, Inc., 76 $9^{\mathrm{th}}$ Avenue, $4^{\mathrm{th}}$ Floor, New York, NY, 10011. \email{muthu@google.com}}
%
%
\maketitle

\abstract{
Modern commercial Internet search engines display advertisements along side the search results in response to user
queries. Such sponsored search relies on market mechanisms to elicit prices for these
advertisements, making use of an
auction among advertisers who bid
in order to have their ads shown for specific
keywords. We present an overview of the current systems for such auctions and 
also describe the underlying game-theoretic aspects. 
The game involves three parties---advertisers, the search engine, and search users---and we present example 
research directions that emphasize the role of each. 
The algorithms for bidding and pricing in these games use techniques from three
mathematical areas: mechanism design, optimization, and statistical estimation. 
Finally, we present some challenges in sponsored search advertising.}

\section{Introduction}


Targeted advertisements on search queries is an increasingly important
advertising medium, attracting large numbers of advertisers and users.
When a user poses a query, the search engine returns search results
together with advertisements that are placed into positions, usually
arranged linearly down the page, top to bottom.  On most major search
engines, the assignment of ads to positions is determined by an
auction among all advertisers who placed a bid on a keyword that
matches the query. The user might click on one or more of the ads, in
which case (in the pay-per-click model\index{Pay-per-click model}) the advertiser receiving the
click pays the search engine a price determined by the auction.

In the past few years, the sponsored search model has been highly
successful commercially, and the research community is attempting to
understand the underlying dynamics, explain the behavior of the market
and improve the auction algorithms.  This survey will provide an overview
of the algorithmic issues in sponsored search. 

The basic view we emphasize is the role of the 
{\em three parties}\index{Three parties in sponsored search}. 
\begin{itemize}
\item
The first party is the {\em advertisers} who 
have multiple objectives in seeking to place advertisements. Some advertisers
want to develop their brand, some seek to make sales, and yet others advertise for
defensive purposes on specific keywords central to their business. 
Some have budget constraints, while others are willing to spend as much as it takes to achieve their goal.
Some seek to obtain many clicks and eyeballs, yet others attempt to optimize
their return on investment.  So, in general, advertisers are of varied types. 

\item
The second party is the {\em auctioneer}, in this case, the search
engine.  The search engines have to balance many needs.  They must
maintain useful search results and have advertisements enhance, rather
than interfere with, the search experience.  They need to make sure
the advertisers get their needs fulfilled, and at the same time ensure
that the market the advertisers participate in is efficient and
conducive to business.

\item
The third party is perhaps the most important in the game: these are {\em search
users}. Users come to search engines for information and pointers. In addition, they also
come to discover shopping opportunities, good deals, and new products. 
There are millions of users with different goals and behavior patterns with
respect to advertisements. 
\end{itemize}

These three parties induce a fairly sophisticated dynamics. While economic
and game theory provide a well-developed framework for understanding the auction game
between the advertisers and the auctioneer, the community has had to generalize such methods
and apply them carefully to understand the currently popular Internet auctions. 
Likewise, while there has been recent work on understanding models of user behavior for
posing search queries and their click behavior for search responses, little is known about
user behavior on advertisements, and crucially, these affect the value of the slots and
thus the very goods that are sold in auction. 

In this survey, we will show examples of research themes in algorithmic, optimization and 
game-theoretic issues in sponsored search. In particular, we present three
examples each emphasizing the perspective of one of the three different parties involved
in sponsored search: the advertisers (who act as the bidders), the
search engine (who acts as the auctioneer), and the search engine user
(who determines the commodity). More specifically,
\begin{itemize}
\item
We present results for how an advertiser should choose their bids given the currently used 
auction mechanism and implicit user behavior models. This result appears as~\cite{FMPS}.
It shows that a very simple bidding strategy is very effective for the advertiser. 

\item
We study a new mechanism for the auctioneer to allocate advertisements
to slots in order to optimize efficiency, and analyze the
game-theoretic aspects of this mechanism. This result appears
in~\cite{FMNP}. It shows that a simple price-setting mechanism is suitable for 
determining the outcome of several auctions simultaneously for the auctioneer. 

\item
We present a novel Markovian model of user behavior when shown advertisements, and for this 
model, develop mechanisms and game theory. This result appears in~\cite{AFMP08}. It shows
that under a model of user behavior more general than the one that is implicit in existing 
auctions, entirely different allocation and pricing will be optimal. Hence, user models have 
significant impact.
\end{itemize}
The results above are joint work with Gagan Aggarwal, Evdokia
Nikolova, Martin P\'al and Cliff Stein, and represent work done at Google Research. 

In the rest of the document, we will first describe the foundations behind the
existing auctions. Then we will describe the three results above. After that, we will be 
able to point to open issues and provide concluding remarks more generally on Internet 
advertising and auctions. 

\section{Existing Auctions}
\label{sec:existing}

\begin{figure}
\begin{center}
\includegraphics[width=4.5in]{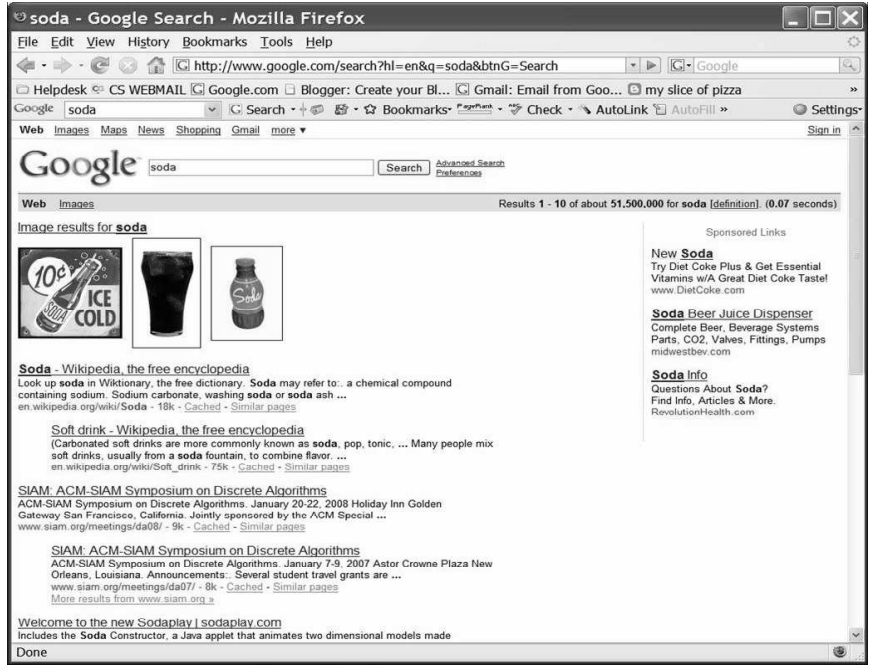}
\end{center}
\caption{Screen shot of a user query with the search results on the left, and the ads on the right.}
\label{fig:scr}
\end{figure}

The basic auction behind sponsored search\index{Sponsored search auction} occurs when a user submits a
query to the search engine. The screen shot in Figure~\ref{fig:scr} shows an example  user query ``soda'' and search results page 
returned from the search engine.
This page includes web search results on the left, and independently, a set of three text ads on the 
right, arranged linearly top to bottom, clearly marked ``Sponsored links.'' 
Each advertiser $i$ has previously submitted a bid $b_i$ stating their
value for a click, tying their bid to a specific {\em keyword}.  The auction is held in real-time among advertisers
whose keywords match that user's query.  The result of the auction
is the list of advertisements on the right. So, after selecting the set of eligible (matching) ads, running the auction involves the search engine 
determining (a) the ordering of bidders and (b) pricing.

\begin{itemize}
\item
{\bf Ordering:} The most natural ordering is to sort by decreasing bid, but
that does not take into account the quality of ads and their
suitability to users.  Thus, it is common practice to place the
bidders in descending order of $b_i \ctr_i$, where $\ctr_i$ is what is
called the {\em click-through-rate} (ctr)\index{Click-through rate (ctr)}\index{ctr (click-through rate)} of advertiser $i$, i.e., the
probability that a user will click on the ad, given that the user
looks at it. (The ctr is usually measured by the
search engine.)  This is the ordering currently in use by search
engines like Yahoo!\index{Yahoo!} and Google.\index{Google}

\item
{\bf Pricing:} The natural method is to make bidders pay what they bid, but
that leads to well-known race conditions~\cite{EOS}. Instead, the most common method is to
use a ``generalized second price'' (GSP) auction\index{GSP (Generalized Second-Price) auction}.  Say the positions
are numbered $1, 2,...$ starting at the top and going down, and the
bidder at position $i$ has bid $b_i$.  In GSP, the price for a click for the
advertiser in position $i$ is determined by the advertisement below it
and is given by $b_{i+1} \ctr_{i+1}/\ctr_i$, which is the minimum they
would have needed to bid to attain their position.
\end{itemize}

\ignore{
{\bf TODO: 
Truthful. GSP is not truthful. VCG auction, definition of Nash equilibirum, envy-free equilibrium, 
separable, etc. Cut and paste from Eyal paper? Will be good to break the following into 
specific theorems.  Jon: I think that we actually don't want to give too much detail
here....  since we're never actually going back to the notion of
profit-maximization, or VCG (except in the context of the user models
stuff, where we just use it as a black box) we can just do this in
english, and cite nisan's book as a good survey.}
}

The first academic treatments of the sponsored search auction were
naturally from the perspective of auction and game theory\index{Auction theory}.  Fixing
this ordering of the bidders, authors in~\cite{EOS,Varian,AGM}
focused on understanding the implications of different pricing
schemes, assuming strategic behavior on the part of the advertisers.
The setting of the game that is modeled in this work is as follows:
each advertiser has a {\em private value} $v_i$ for a click from this
user, and wants to set a bid that maximizes her {\em utility} $u_i$.
The natural economic utility model in this context would be profit:\index{Profit utility model}
i.e., $$u_i = (v_i - p_i) c_i,$$ where $p_i$ is the price per click, and
$c_i$ is the probability of a click occurring.  Of course $c_i$ is
determined by the {\em user}, and may depend on any number of factors.
One common model is to assume that the click probability is {\em
separable}~\cite{AGM}:\index{Separable click probability} if ad $i$ is placed into position $j$, then $c_i = \ctr_i
\pn_j$ where $\ctr_i$ is the ad-specific ``click-through rate'' and
$\pn_j$ is a position-specific visibility factor.\index{Position-specific visibility factor} (We will later
explore other utility models in Sections~\ref{sec:uniform}
and~\ref{sec:scheduling} when we incorporate budgets, as well as
non-separable user models in Section~\ref{sec:markov}.)

Natural questions in this context include asking whether there is a
{\em pure-strategy\index{Pure-strategy Nash equilibrium} Nash equilibrium\index{Nash Equilibrium}} of this game, and analyzing the
economic efficiency and revenue of such equilibria.  By {\em economic
efficiency}\index{Economic efficiency} we mean the total advertiser value generated by the
assignment.  This is also commonly referred to as the {\em social
welfare}.\index{Social welfare}  In the context of sponsored search, the efficiency is the
sum of the individual advertisers' values; i.e., $\sum_i c_i v_i$,
where $c_i$ is the probability that $i$ will receive a click under
this assignment and $v_i$ is $i$'s private value for a click.  By a
{\em pure-strategy Nash equilibrium} we mean a set of bids such that
no single bidder can change her bid and increase her utility.

Among the most desirable properties of a mechanism is to be {\em
truthful},\index{Truthful mechanism} which is also referred to as being {\em incentive
compatible}.\index{Incentive compatible}  This property says that each bidder's best strategy,
regardless of the actions of other bidders, is simply to report her
true value; i.e., submit $v_i$ as her bid.  Truthfulness immediately
implies the existence of a pure-strategy Nash equilibrium (where every
bidder reports $v_i$).  Furthermore, it is simple to compute economic
efficiency, since the assignment (and thus the efficiency) is simply a
function of the values $v_i$.  Unfortunately, it turns out that the
GSP auction is {\em not} truthful.  However, there is a
pricing scheme that is truthful, which is based on an application of
the famous Vickrey-Clarke-Groves (VCG) mechanism~\cite{V,C,G}\index{VCG (Vickrey-Clarke-Groves) mechanism}.
Furthermore, the GSP auction, while not truthful, still has a
well-understood pure-strategy Nash equilibrium:

\begin{theorem}[\cite{EOS,Varian,AGM}]
\label{thm:classic}
Suppose we have a set of bidders participating in a particular
sponsored search auction.  
Assume each bidder has a private value and a profit-maximizing utility function.
Suppose further that the click probabilities are separable. 
Then, the GSP auction is not truthful, but it does have a pure-strategy Nash equilibrium
whose outcome (in terms of assignment and prices) is equivalent to an application of the truthful VCG auction.
\end{theorem}

For a more detailed discussion of this line of research, we refer the reader to~\cite{nisan_chapter}.
Authors in~\cite{AGM} also show that under a more general click
probability model, there is a pricing method that is truthful. (This
pricing method reduces to the VCG pricing method when the
click-through rates are separable.)  Furthermore, they show that in
this more general setting the GSP has a Nash equilibrium that has the
same outcome as their mechanism.


\subsection{Practical Aspects}
The results described above regard GSP as an isolated auction,
abstracting away the context of the larger system of which it is a
part.  While this is useful from a modeling perspective, there are
many other elements that make sponsored search a more complex
environment.  Here we list some of those complicating factors, and
mention examples of work done to address them.
\begin{itemize}
\item
{\bf Multiple queries, multiple keywords.} Each sponsored auction is
conducted for a particular search engine user with a potentially
unique query.  There are perhaps millions of such queries every day.
Advertisers must submit bids on {\em keywords},\index{Keywords in sponsored search} and cannot adjust
those bids on a per-query basis.  The degree to which the keyword
matches a particular query determines not only whether the advertiser
will participate in the auction (and also who her competitors will
be), but also can factor into the click-through rate $\ctr_i$ that is
used for ranking.  Theorem~\ref{thm:classic} only applies to the case
where the same auction---with the same set of advertisers, and the
same click-through rates---is repeated, and the bids qualify only for
that set of auctions.  A lot of the work mentioned below takes on this
complication in various ways; we give two such examples in
Sections~\ref{sec:uniform} and~\ref{sec:scheduling}.

\item
{\bf Budgets.}  In the private-value model\index{Private-value model} each advertiser has a value
$v_i$ per click, but is willing to spend an arbitrary amount to
maximize her profit.  In reality, many advertisers have operating
budgets or spending targets, and simply want to maximize their value
given the constraints of that budget.  This budget can be reported to
the search engine, who can then employ techniques to use the budget
efficiently.  Analysis of incentives becomes more difficult in
the presence of budgets.  This has been addressed 
in e.g.,~\cite{BCIMS,MS,MNS,MSVV,MNS,AMT,mps,RW,FMPS,AFMP08}, and
 we discuss two examples in much more detail in
Sections~\ref{sec:uniform} and~\ref{sec:scheduling}.  

\item
{\bf Reserve prices.}  The major search engines enforce {\em reserve
prices},\index{Reserve prices in sponsored search} dictating the minimum price that an advertiser can pay for a
click. Sometimes these reserve prices will even be specific to a
particular bidder.  Reserve prices are useful for controlling quality
on the search results page, and also have implications for revenue.
The effect of reserve prices on the game theory of sponsored search is
discussed in detail in~\cite{EFMM08}.

\item
{\bf Interdependent click probabilities.}  The ``separable''
assumption implies that an advertiser's click probability depends only
on the properties and position of her own ad.  This ignores the other
ads on the search results page, which certainly affect the user
experience, and therefore the click probability of this advertiser.
We discuss this further in Section~\ref{sec:markov}.

\item
{\bf Branding.}  The private click-value model assumes that a click is
what the advertiser is ultimately interested in.  However a {\em
branding} advertiser could be interested in her ad appearing in a high
position, but not really care whether or not it gets a click (other
than due to the fact that they only pay if it does). (Indeed, a recent
empirical study by the Interactive Advertising Bureau and
Nielsen//NetRatings concluded that higher ad positions in paid search
have a significant brand awareness effect~\cite{iabstudy}.)  Thus we
might be interested in an auction where an advertiser can express the
lowest position she is willing to tolerate for her ad.  This is the
approach taken in~\cite{afm}, where Theorem~\ref{thm:classic} is
generalized to this setting.

\item
{\bf Conversions.}  The private click-value model also assumes that
each click is worth the same to an advertiser, which is not always the
case in practice.  Indeed many advertisers track whether or not a
click leads to a {\em conversion},\index{Conversions in sponsored search} which is some sort of event on the
linked page (e.g., a sale, a sign-up, etc.) Given this data, the
advertiser can learn which keywords lead to conversions and therefore
which clicks are worth more to them.

\item
{\bf Estimating various parameters.}  Most work in the context of the game
theory of sponsored search has assumed that the parameters like
click-through rate and position visibility are known.  However,
estimating these parameters is a difficult task (e.g.,~\cite{IJMT,RDR}).
Indeed, there is an inherent tradeoff between learning these
parameters and applying them; one cannot learn that an ad has a bad
ctr unless it is exposed to the user, but then it was a bad idea to
show it in the first place.  This ``exploration/exploitation''
tradeoff turns out to be related to the ``multi-armed bandit''\index{Multi-armed bandit problem}
problem (see e.g.~\cite{GP,WVLL,EMM02}).

\item
{\bf Incomplete Knowledge.}  Both the advertisers and the search
engine have incomplete knowledge of the ``inventory'' available to
them, since they do not know which queries will arrive.  In addition
the bidders do not know the other bids or click-through rates.  This
makes the advertiser's optimization problem much more difficult (see
e.g., \cite{Lahaie,BCEIJM,CDEGHKMS,FMPS,VR,LQ,ZL07,SL,RW}).  From the search
engine's point of view, we can model incomplete knowledge of the
future as an {\em online algorithm}; see e.g.~\cite{MSVV,RW,mps,WVLL,MNS,MS,AG,GP,GM07,GM08}.
\end{itemize}

\section{The Advertiser's Point of View: Budget Optimization}
\label{sec:uniform}

The perspective in this section is the advertisers.  The
challenge from an advertiser's point of view is to understand and
interact with the auction mechanism.  The advertiser determines a set
of keywords of their interest\footnote{The choice of keywords
is related to the domain-knowledge of the advertiser, user behavior
and strategic considerations. Internet search companies provide
the advertisers with summaries of the query traffic which is useful for them to 
optimize their keyword choices interactively.
We do not directly address the choice of keywords in this section, which is
addressed elsewhere~\cite{RW}.} and then must create ads, set the bids
for each keyword, and provide a total (often daily) budget.

While the effect of
an ad campaign in any medium is a sophisticated phenomenon that is difficult to quantify, one commonly accepted (and easily quantified) notion
in search-based advertising on the Internet is to {\em maximize the number of clicks}.  The 
Internet search companies are supportive towards advertisers and 
provide statistics about the history of click volumes and prediction 
about the future performance of various keywords. Still, this is a sophisticated
problem for the following reasons (among others):
\begin{itemize}
\item
Individual keywords have significantly different 
characteristics from each other; e.g., while ``fishing'' is a broad keyword 
that matches many user queries and has many competing advertisers,
``humane fishing bait'' is a niche keyword that matches only a few queries, but
might have less competition. 
\item
There are complex {\em interactions} between keywords because a user
query may match two or more keywords, since the advertiser
is trying to cover all the possible keywords in some domain. In effect
the advertiser ends up competing with herself.
\end{itemize}
As a result, the advertisers face a challenging  
optimization problem. 
The focus of the work in~\cite{FMPS} is to solve this optimization problem.

\heading{Problem Formulation.}
We present a short discussion and formulation of the 
optimization problem faced by advertisers; a more detailed 
description is in Section~\ref{sec:model}. 

A given advertiser sees the state of the auctions for search-based 
advertising as follows. There is a set $\K$ of keywords of interest;
in practice, even small advertisers typically have a large set $\K$.
There is a set $\Q$ of queries posed by the users. 
For each query $q \in Q$, there are functions giving the $\cl_q(b)$ and
$\cost_q(b)$ that result from bidding a particular amount $b$ in the auction for
that query, which we will see a more formal model of in the next
section.
There is a bipartite graph $\G$ on the two vertex sets representing $\K$
and $\Q$.  For any query $q \in \Q$, the neighbors of $q$ in $\K$ are the keywords that are said to ``match'' the query
$q$.\footnote{The particulars of the matching rule are determined by
the Internet search company; here we treat the function as arbitrary.}
 
The {\em budget optimization problem} is as follows.  Given graph $\G$
together with the functions $\cl_q(\cdot)$ and $\cost_q(\cdot)$ on the
queries, as well as a budget $\B$, determine the bids
$b_k$ for each keyword $k \in \K$ such that $\sum_q \cl_q(b_q)$ is
maximized subject to $\sum_q \cost_q(b_q) \leq \B$, where the
``effective bid'' $b_q$ on a query is some function of the keyword bids in the
neighborhood of $q$.

While we can cast this problem as a traditional optimization problem,
there are different challenges in practice depending on the
advertiser's access to the query and graph information, and indeed the reliability of
this information (e.g., it could be based on unstable historical
data).  Thus it is important to find solutions to this problem that
not only get many clicks, but are also simple, robust and less reliant on 
the information.  The notion of a ``uniform'' strategy is defined in~\cite{FMPS} which is
essentially a strategy that bids uniformly on all keywords.  Since
this type of strategy obviates the need to know anything about the
particulars of the graph, and effectively aggregates the click and
cost functions on the queries, it is quite robust, and thus desirable in
practice.  What is surprising is that uniform strategy actually performs well, which is proved in~\cite{FMPS}.

\heading{Main Results and Technical Overview.}
Some positive and negative results are given in~\cite{FMPS} for the  budget optimization problem:

\begin{itemize}
\item
Nearly all formulations of the problem are NP-Hard. 
In cases slightly more general than the 
formulation above, where the clicks have weights, the problem is
inapproximable better than a factor of $1 - \frac 1 e$, unless P=NP.
\item
There is a $(1-1/e)$-approximation algorithm for the budget
optimization problem.  The strategy found by the algorithm is a {\em
two-bid uniform strategy}, which means that it randomizes between
bidding some value $b_1$ on all keywords, and bidding some other value
$b_2$ on all keywords until the budget is exhausted\footnote{This type of strategy can also be
interpreted as bidding one value (on all keywords) for part of the
day, and a different value for the rest of the day.}.
This approximation ratio is tight for uniform strategies.
There is also a
$(1/2)$-approximation algorithm that offers a {\em single-bid uniform
strategy}, only using one value $b_1$. (This is tight for single-bid uniform strategies.)  These strategies can be
computed in time nearly linear in $|Q|+|K|$, the input size.
\end{itemize}

Uniform strategies may appear to be naive in first consideration
because the keywords vary significantly in their click and cost functions, and there may be
complex interaction between them when multiple keywords are
relevant to a query. After all, the optimum can configure arbitrary bids on each
of the keywords. Even for the simple case when the graph is a {\em matching}, the optimal algorithm 
involves placing different bids on different keywords via a knapsack-like packing
(Section~\ref{sec:model}). 
So, it might be surprising that a simple two-bid uniform strategy is $63\%$ or more effective compared
to the optimum.
In fact, our proof is stronger, showing that this strategy is $63\%$ effective 
against a strictly more powerful adversary who can bid independently on the 
{\em individual queries}, i.e., not be constrained by the interaction imposed by 
the graph $\G$. 

We will also look at the simulations conducted in~\cite{FMPS} using
real auction data from Google.  The results of these simulations
suggest that uniform bidding strategies could be useful in practice.
However, important questions remain about (among other things)
alternate bidding goals, on-line or stochastic bidding
models~\cite{mps}, and game-theoretic concerns~\cite{BCIMS}, which we
briefly discuss in Section~\ref{sec:conc}.




\subsection{Modeling a Keyword Auction}
\label{sec:model}

We begin by considering the case of a {\em single} keyword that
matches a {\em single} user query.  In this section we define the
notion of a ``query landscape'' that describes the relationship
between the advertiser's bid and what will happen on this query as a
result of this bid~\cite{autobidder}.  This definition will be central to the discussion
as we continue to more general cases.

The search results page for a query contains 
$\poss$ possible positions in which our ad can appear.
We denote the highest (most
favorable) position by $1$ and lowest by $\poss$.  
Assuming a separable user model, associated with each position $i$ is a value $\bctr[i]$ that denotes
the click probability if the ad appears in position $i$.\footnote{We
leave out the ad-specific factor $\ctr_i$ from this section for
clarity, but all the results in~\cite{FMPS} generalize to this case as
well.}   We
assume throughout this section that that $\bctr[i] \leq \bctr[j]$ if
$j<i$, that is, higher positions receive at least as many clicks as
lower positions.

In order to place an ad on this page, we must enter the {\em GSP
auction} that is carried out among all advertisers that have submitted
a bid on a keyword that matches the user's query. We will refer to
such an auction as a {\em query auction},\index{Query auction} to emphasize that there is
an auction for each query rather than for each keyword.  In GSP, the
advertisers are ranked in decreasing order of bid, and each advertiser
is assigned a price equal to the amount bid by the advertiser below
them in the ranking.
Let $(\bid[1], \dots, \bid[\poss])$ denote the bids of the top $\poss$
advertisers in this query auction.  For notational convenience, we
assume that $\bid[0] = \infty$ and $\bid[{\poss}] = \bctr[\poss] =0$.
Since the auction is a generalized second price auction, higher bids
win higher positions; i.e. $b[i] \geq b[{i+1}]$.  
Suppose
that we bid $\bid$ on some keyword that matches the user's query, then
our position is defined by the largest $\bid[i]$ that is at most
$\bid$, that is, 
\begin{equation}
\label{eq:posdef}
\iposs(\bid) = \argmax_i(\bid[i]: \bid[i] \leq \bid) .
\end{equation}
Since we 
only pay if the user clicks (and that happens with probability
$\bctr[i]$), our expected {\em cost} for winning position $i$ would be
$\cost[i] = \bctr[i] \cdot \bid[i],   \mbox{ where } i = \iposs(\bid) .$
We use $\cost_q(b)$ and $\cl_q(b)$ to denote the expected cost and
  clicks that result from having a bid $b$ that qualifies for a query  auction
$q$, and thus
\begin{equation} \cost_q(b) = \bctr[i] \cdot \bid[i] \;\;\; \mbox{ where  }  i = \iposs(\bid)
, 
\label{eq:cost1}
\end{equation}
\begin{equation}
\cl_q(b) = \bctr[i]\;\;\; \mbox{ where  }  i = \iposs(\bid) . 
\label{eq:click1}
\end{equation}

When the context is clear, we drop the subscript $q$.  The following observations about cost and clicks follow immediately from the definitions and
equations~\eqref{eq:posdef}, \eqref{eq:cost1} and~\eqref{eq:click1}.
We use $\reals_+$ to denote the nonnegative reals.
\begin{proposition}
\label{observations}
For $b \in \reals_+$,
\begin{enumerate} 
\item  The tuple ($\cost_q(b),\cl_q(b)$) can only take on one of a
finite set of values $\lan_q = \{ (\cost[1], \bctr[1]), \dots, (\cost[\poss],
\bctr[\poss]) \}$.
\item Both $\cost_q(b)$ and $\cl_q(b)$ are non-decreasing functions of
$b$.  
\item Cost-per-click (cpc)\index{cpc (Cost-Per-Click)}\index{Cost-per-click (cpc)} $\cost_q(b) / \cl_q(b)$ is non-decreasing in $b$, and is always at most the bid; i.e., $\cost_q(b) / \cl_q(b) \leq b$.
\end{enumerate}
\label{obs:simple}
\end{proposition}

\subsubsubsection{Query Landscapes}\index{Query landscape}
We can summarize the data contained in the functions $\cost(b)$ and
$\cl(b)$ as a collection of points in a plot of cost vs. clicks,
which we refer to as a {\em landscape}.\index{Landscape in a query auction}  For example, for a query with four slots, a landscape might look like Table
\ref{table:landscape-ex}.

\begin{figure}[h]
\begin{center}
\includegraphics[height=2.0in]{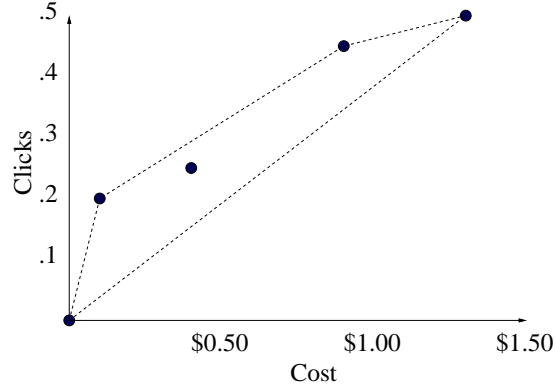}
\end{center}
\caption{A bid landscape.}
\label{fig:landscape1}
\end{figure}

\begin{table}[h]
\begin{center}
\begin{tabular}{r|r|r|r}
bid range           & cost per click & cost  & clicks\\ \hline
~ [\$2.60, $\infty$) & \$2.60         & \$1.30& .5    \\
~ [\$2.00, \$2.60)   & \$2.00         & \$0.90& .45   \\
~ [\$1.60, \$2.00)   & \$1.60         & \$0.40& .25   \\
~ [\$0.50, \$1.60)   & \$0.50         & \$0.10& .2   \\
~ [\$0, \$0.50)      & \$0            & \$0   & 0        
\end{tabular}
\caption{A {\em landscape} for a query}
\label{table:landscape-ex}
\end{center}
\end{table}
It is convenient to represent this data graphically as in
Figure~\ref{fig:landscape1} (ignore the dashed line for now).  Here we graph clicks as a function of
cost.  Observe that in this graph, the cpc $(\cost(b) / \cl(b))$ of each
point is the reciprocal of the slope of the line from the origin to
the point.  Since $\cost(b)$, $\cl(b)$ and $\cost(b) / \cl(b)$ are
non-decreasing, the slope of the line from the origin to successive points
on the plot decreases.  This condition is  slightly weaker
than concavity.  

Suppose we would like to solve the budget optimization problem for a single
query landscape.\footnote{Of course it is a bit unrealistic to
imagine that an advertiser would have to worry about a budget if only
one user query was being considered; however one could imagine
multiple instances of the same query and the problem scales.}  
As we increase our bid from zero, our cost increases and our expected number of clicks increases, 
and so we simply submit the highest bid such that we remain within our budget.

One problem we see right away is that since there are only a finite
set of points in this landscape, we may not be able to target
arbitrary budgets efficiently.  Suppose in the example from
Table~\ref{table:landscape-ex} and Figure~\ref{fig:landscape1} that we had a budget of $\$1.00$.
Bidding between $\$2.00$ and $\$2.60$ uses only $\$0.90$, and so we
are under-spending.  Bidding more than $\$2.60$ is not an option, since
we would then incur a cost of $\$1.30$ and overspend our budget.

\subsubsubsection{Randomized strategies}
To rectify this problem and better utilize our available budget, we
allow {\em randomized bidding strategies.}\index{Randomized bidding strategy}  Let 
$\bidd$ be a distribution on bids $\bid \in \reals_+$.
Now we define
$\cost(\bidd) = E_{\bid \sim \bidd} [\cost(\bid)]$ and $\cl(\bidd) =
E_{\bid \sim \bidd} [\cl(\bid)]$.  
Graphically, the possible values of
$(\cost(\bidd), \cl(\bidd))$ lie in the convex hull\index{Convex hull} of the landscape
points. 
This is represented in 
Figure~\ref{fig:landscape1} by the dashed line.


To find a bid distribution $\bidd$ that maximizes
clicks subject to a budget, we simply draw a vertical line on the plot
where the cost is equal to the budget, and find the highest point on this line in
the convex hull.  This point will always be the convex
combination of at most {\em two} original landscape points which
themselves lie {\em on} the convex hull.  Thus, given the point on the
convex hull, it is easy to compute a distribution on two bids
which led to this point.  Summarizing,

\begin{lemma}
\rm{\cite{FMPS}}
\label{lem:convex}
If an advertiser is bidding on one query, subject to a budget $\B$,
then the optimal strategy is to pick a convex combination of (at most)
two bids which are at the endpoints of the line on the convex hull at
the highest point for cost $\B$. 
\end{lemma}

There is one subtlety in this formulation.  Given any bidding strategy,
randomized or otherwise, the resulting cost is itself a random
variable representing the expected cost.  Thus if our budget
constraint is a hard budget, we have to deal with the difficulties
that arise if our strategy would be over budget.
Therefore, we think of our budget constraint as {\em soft},\index{Soft budget constraint} that is, we
only require that our expected cost be less than the budget.  
In practice, 
the budget is often an average daily budget, and
thus we don't worry if we exceed it one day, as long as we are
meeting the budget in expectation.  Further, 
either the advertiser or the search engine (possibly both), monitor the
cost incurred over the day; hence, the advertiser's bid can be changed to
zero for part of the day, so that the budget is not overspent.\footnote{See \url{https://adwords.google.com/support/bin/answer.py?answer=22183}, for example.}  
Thus in the remainder of this section, we will formulate  a budget constraint
that only needs to be respected in expectation.

\subsubsubsection{Multiple Queries: a Knapsack Problem}
\label{sec:knapsack} 
As a warm-up, we will consider next the case when we have a set of
queries, each with its own landscape\index{Knapsack problem}.  We want to bid on each
query independently subject to our budget:
the resulting optimization problem is
a small generalization of the {\em fractional knapsack}\index{Fractional knapsack problem}
problem, and was solved in~\cite{autobidder}. 


The first step of the algorithm is to take the convex hull of each
landscape, as in Figure~\ref{fig:landscape1}, and remove any
landscape points not on the convex hull. 
Each
piecewise linear section of the curve
represents the
incremental number of clicks and cost incurred by moving one's bid 
from one particular value 
to another.   
We regard these ``pieces'' as {\em items} in an instance of fractional
knapsack with {\em value} equal to the incremental number of clicks and {\em size}
equal to the incremental cost.
More precisely, 
for each piece connecting two
consecutive bids $b'$ and $b''$ on the convex hull, we create a knapsack item with 
value $[\cl(b'') - \cl(b')]$ and 
size $[\cost(b'') - \cost(b')]$.
  We then emulate the greedy algorithm for
knapsack, sorting by value/size (cost-per-click), and choosing greedily until the
budget is exhausted.  

In this reduction to knapsack we have ignored the fact that some of
the pieces come from the same landscape and cannot be treated
independently.  However, since each curve is concave, the pieces that
come from a particular query curve are in increasing order of
cost-per-click; thus from each landscape we have chosen for our
``knapsack'' a set of pieces that form a prefix of the curve.

\subsubsection{Keyword Interaction}\index{Keyword interaction}

In reality, search advertisers can bid on a
large set of keywords, each of them qualifying for a different (possibly 
overlapping) set of queries, but
most search engines do not allow an
advertiser to appear twice in the same search results
page.\footnote{See \url{https://adwords.google.com/support/bin/answer.py?answer=14179}, for example.}  
Thus, if an
advertiser has a bid on two different keywords that match the same
query, this conflict must be resolved somehow.  For example, if an
advertiser has a bid out on the keywords ``shoes'' and ``high-heel,''
then if a user issues the query ``high-heel shoes,'' it will match on
two different keywords.  The search engine specifies, in advance, a
rule for resolution based on the query the keyword and the bid.  A
natural rule is to take the keyword with the highest bid, which we adopt here, but
our results apply to other resolution rules. 

We model the keyword interaction problem using an undirected bipartite graph $\G =
(\K \cup \Q, \edges)$ where $\K$ is a set of keywords and $\Q$ is a set of
queries. Each $q \in \Q$ has an associated landscape, as defined by 
$\cost_q(\bid)$ and $\cl_q(\bid)$.
An edge $(k,q) \in \edges$ means that keyword $k$
matches query $q$.

The advertiser can control their individual {\em keyword bid vector} $\bidv
\in \reals_+^{|K|}$ specifying a bid $\bidv_k$ for each keyword $k \in \K$.  
(For now, we do not consider randomized bids, but we will introduce
that shortly.)  Given a particular bid vector $\bidv$ on the keywords,
we use the resolution rule of taking the maximum to define the ``effective
bid'' on query $q$ as
$$\bid_q(\bidv) = \max_{k: (k,q) \in \edges} \bidv_k. $$
By submitting a bid vector $\bidv$, the advertiser receives some
number of clicks and pays some 
cost on each
keyword.  
We use the term {\em spend} to denote the total cost; 
similarly, we use the term {\em traffic} to
denote the total number of clicks:
\begin{eqnarray*}
\tcost(\bidv) \!=\! \sum_{q \in Q} \cost_q(\bid_q(\bidv));~~~
\tcl(\bidv)  \!=\! \sum_{q \in Q} \cl_q(\bid_q(\bidv))
\end{eqnarray*}

\noindent We also allow randomized strategies, where an advertiser gives a
distribution $\bidvd$ over bid vectors $\bidv \in \reals_+^{|K|}$.
The resulting spend and traffic are given by 
\begin{eqnarray*}
\tcost(\bidvd) \!=\! E_{\bidv \sim \bidvd} [\tcost(\bidv)];~~~\tcl(\bidvd) \!=\! E_{\bidv \sim \bidvd} [\tcl(\bidv)]
\end{eqnarray*}
We can now state the problem in its full generality:\index{Budget Optimization Problem}

\medskip
\begin{center}
\noindent
\begin{svgraybox}
\underline{\kbidding}  \\
\noindent {\bf Input:} a budget $\B$, a keyword-query graph $\G
= (\K \cup \Q, \edges)$, and landscapes $(\cost_q(\cdot),\cl_q(\cdot))$ for
each $q \in \Q$.   \\
\noindent {\bf Find:} a distribution $\bidvd$ over bid vectors $\bidv
\in \reals_+^{|K|}$ such that $\tcost(\bidvd) \leq \B$ and
$\tcl(\bidvd)$ is maximized.
\end{svgraybox}
\end{center}
\medskip

We conclude this section with a small example to illustrate some
feature of the budget optimization 
problem. Suppose
you have two keywords $K = \set{u, v}$ and two queries
$Q=\set{x,y}$ and edges  $E= \set{(u,x), (u,y),
(v,y)}.$  
Suppose query $x$ has one position with ctr $\bctr^x[1] = 1.0$, and there is one bid $b^x_1 = \$1$.  
Query $y$ has two positions with ctrs $\bctr^y[1] = \bctr^y[2] = 1.0$, and bids $b^y_1 = \$\eps$ and $b^y_2 = \$1$
To get any clicks from $x$, an advertiser must bid at least $\$1$ on
$u$.  However, because of the structure of the graph, if the
advertiser sets $b_{u}$ to \$1, then his effective bid is $\$1$ on both
$x$ {\em and} $y$.  Thus he must trade-off between getting the clicks
from $x$ and getting the bargain of a click for $\$\eps$ that would be
possible otherwise.

\subsection{Uniform Bidding Strategies}

As shown in~\cite{FMPS}, solving the \kbidding problem in its full generality is difficult.  
In addition, it may be difficult to reason about strategies that involve
arbitrary distributions over arbitrary bid vectors.  
Advertisers generally prefer strategies that are easy to
understand, evaluate and use within their larger goals.
With this motivation, we look at restricted classes of strategies that
we can easily compute, explain and analyze.


We define a {\em uniform  bidding strategy} to be a distribution
$\bidvd$ over bid vectors $\bidv \in \reals_+^{|K|}$ where each bid
vector in the distribution is of the form $(b, b, \dots, b)$ for some real-valued bid
$b$.  In other words, each vector in the distribution bids the same
value on every keyword.\index{Uniform bidding strategy}

Uniform strategies have several advantages.
First, they do not depend on the edges of the
interaction graph, since all effective bids on queries are the same.
Thus, they are effective in the face of limited or noisy information 
about the keyword interaction graph.
Second, uniform strategies are also independent of the priority rule being
used.  Third, any algorithm that gives an approximation guarantee will then be valid for {\em any} interaction
graph over those keywords and queries.

Define a {\em two-bid strategy}\index{Two-bid strategy} to be a uniform strategy which puts
non-zero weight on at most two bid vectors.  Given the landscapes for
all the queries, we can compute the best uniform strategy in linear
time; the proof also directly implies that there is always an optimal
two-bid strategy:
\begin{lemma}
\rm{\cite{FMPS}}
\label{lemma:ag}
Given an instance of \kbidding in which there are a total of $\tpoints$ points in
all the landscapes, we can find the best uniform strategy in $O(\tpoints \log \tpoints)$ time.
Furthermore, this strategy will always be a two-bid strategy. 
\end{lemma}

The authors in~\cite{FMPS} also consider {\em single-bid} strategies,\index{Single-bid strategy} which are uniform
strategies that put non-zero weight on at most one {\em non-zero}
vector, i.e.  advertiser randomizes between bidding a certain amount
$b^*$ on all keywords, and not bidding at all.  A single-bid strategy
is even easier to implement in practice than a two-bid strategy. For
example, the search engines often allow advertisers to set a maximum
daily budget. In this case, the advertiser would simply bid $b^*$
until her budget runs out, and the ad serving system would remove her
from all subsequent auctions until the end of the day. One could also
use an ``ad scheduling''\index{Ad scheduling} tool offered by some search
companies\footnote{See
\url{https://adwords.google.com/support/bin/answer.py?answer=33227},
for example.} to implement this strategy.  The best single-bid
strategy can also be computed easily from the aggregate landscape.
The optimal strategy for a budget $U$ will either be the point $x$
s.t. $\cost(x)$ is as large as possible without exceeding $U$, or a
convex combination of zero and the point $y$, where $\cost(y)$ is as
small as possible while larger than $U$.

\heading{Approximation Guarantees of Uniform Strategies.}
\label{sec:approx}
In fact, not only are uniform strategies easy to optimize over, they
are also guaranteed to have good performance compared to the optimal
solution.  In the case of single-bid strategies, we have the following:
\begin{theorem}
\label{thm:2approx}
\rm{\cite{FMPS}}
There always exists a uniform single-bid strategy that is $\half$-optimal.
Furthermore, for any $\eps > 0$ there exists an instance for which all single-bid
strategies are at most $(\half + \eps)$-optimal.
\end{theorem}

For general uniform strategies---where a two-bid strategy is always
optimal---\cite{FMPS} proves a tighter approximation ratio:
\begin{theorem}
\label{thm:eapprox}
\rm{\cite{FMPS}}
There always exists a uniform bidding strategy that is $(1-\frac 1
e)$-optimal.  Furthermore, for any $\epsilon>0$, there exists an
instance for which all uniform strategies are at most
$(1-\frac 1 e + \epsilon)$-optimal.
\end{theorem}

Thus if given full information about the landscapes, a bidder has an
efficient strategy to get a large fraction of the available clicks at
her budget.  But perhaps more importantly, these theorems show that
the simple uniform bidding heuristic can perform well.

\subsection{Experimental Results}
\begin{figure}
\begin{center}
\psfrag{#0#}{$0$}
\psfrag{#half#}{$0.5$}
\psfrag{#1#}{$1$}
\psfrag{#Budget#}{Budget}
\psfrag{#Clicks#}{Clicks}
\includegraphics[height=2.35in]{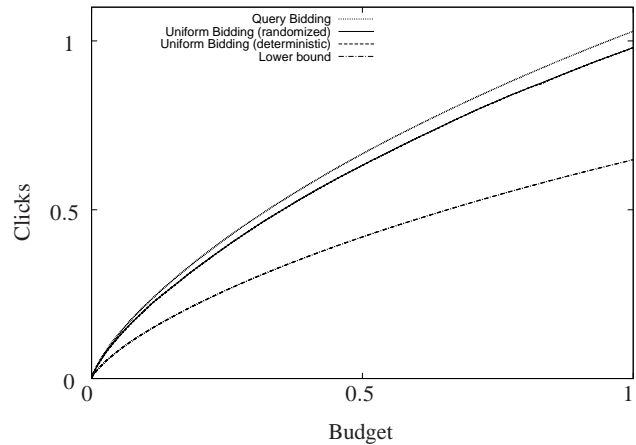}
\end{center}
\caption{An example with real data.}
\label{fig:real}
\end{figure}

The authors in~\cite{FMPS} ran simulations using the data available at Google\index{Google} which we
briefly summarize here.  They took a large advertising campaign, and,
using the set of keywords in the campaign, computed three different
curves (see Figure~\ref{fig:real}) for three different bidding
strategies.  The x-axis is the budget (units removed), and the y-axis
is the number of clicks obtained (again without units) by the optimal bid(s) under each
respective strategy. ``Query bidding'' represents the (unachievable) upper bound $\OPT$,
bidding on each query independently. The ``uniform bidding'' curves
represent the results of applying the
algorithm: ``deterministic'' uses a single bid level, 
while ``randomized'' uses a distribution.
For reference, we include the lower bound of 
a $(e-1)/e$ fraction of the top
curve.

The data clearly demonstrate that the best single uniform bid obtains
almost all the possible clicks in practice.  Of course in a more
realistic environment without full knowledge, it is not always
possible to find the best such bid, so further investigation
is required to make this approach useful.  However, just knowing that there is such a bid available
should make the on-line versions of the problem simpler.

\subsection{Extensions}
\label{sec:conc}

The algorithmic result presented here gives an intriguing heuristic in practice:
bid a single value $b$ on all keywords; at the end of the day, if the
budget is under-spent, adjust $b$ to be higher; if budget is overspent,
adjust $b$ to be lower; else, maintain $b$.  If the scenario does not
change from day to day, this simple strategy will have the same
theoretical properties as the one-bid strategy, and in practice,
is likely to be much better. Of course the scenario does change, however,
and so coming up with a ``stochastic'' bidding strategy\index{Stochastic bidding strategy}
remains an important open direction, explored somewhat
by~\cite{mps,RW}.

Another interesting generalization is to consider weights on the
clicks, which is a way to model {\em conversions}. (A conversion
corresponds to an action on the part of the user who clicked through
to the advertiser site; e.g., a sale or an account sign-up.)  Finally,
we have looked at this system as a black box returning clicks as a
function of bid, whereas in reality it is a complex repeated game
involving multiple advertisers.  In~\cite{BCIMS}, it was shown that
when a set of advertisers use a strategy similar to the one suggested
in~\cite{FMPS}, under a slightly modified first-price auction, the prices
approach a well-understood market equilibrium.

\section{The Search Engine's Point of View: Offline Slot Scheduling}
\label{sec:scheduling}

In the previous section we saw that when we take the GSP auction as
given, and view the world through the lens of the bidder, the
practical problem becomes more complex than what the individual auction was
designed for.  But we could take the question back to the search
engine and ask if there is a more general mechanism that regards the
entire day's worth of queries as part of a single overall game.  This question is addressed in~\cite{FMNP}, where the {\em
Offline Ad Slot Scheduling} problem is defined: given a set of bidders
with bids (per click) and budgets (per day), and a set of slots over
the entire day where we know the expected number of clicks in each
slot, find a schedule that places bidders into slots.  The schedule
must not place a bidder into two different slots at the same time.  In
addition, we must find a price for each bidder that does not exceed
the bidder's budget constraint, nor their per-click bid.  (See below
for a formal statement of the problem.)

A good algorithm for this problem will have high revenue.  Also, we
would like the algorithm to be {\em truthful}; i.e., each bidder will
be incented to report her true bid and budget.  In order to prove
something like this, we need a {\em utility function} for the bidder
that captures the degree to which she is happy with her allocation.
Natural models in this context (with clicks, bids and budgets) are\index{Click maximization utility function}
{\em click-maximization}---where she wishes to maximize her number of
clicks subject to her personal bid and budget constraints, or {\em
profit-maximization}---where she wishes to maximize her profit (clicks
$\times$ profit per click).  The work in~\cite{FMNP} is focused on click-maximization.\footnote{This choice is in part motivated by the presence
of budgets, which have a natural interpretation in this application:
if an overall advertising campaign allocates a fixed portion of its
budget to online media, then the agent responsible for that budget is
incented to spend the entire budget to maximize exposure.  In
contrast, under the profit-maximizing utility, a weak motivation for budgets is a limit on liquidity.
Also, this choice of utility function is out of
analytical necessity: Borgs et al.~\cite{BCIMS} show that under some
reasonable assumptions, truthful mechanisms are impossible under a
profit-maximizing utility.}

We present the efficient mechanism of~\cite{FMNP} for {\em Offline Ad
Slot Scheduling}, which is truthful under click-maximization.
Also, the revenue-optimal mechanism for {\em Offline Ad Slot
Scheduling} is not truthful, but has a Nash equilibrium (under the
same utility model) whose outcome is equivalent to the~\cite{FMNP}
mechanism; this result is strong evidence that the mechanism has
desirable revenue properties.

\heading{Problem Definition.}
The {\em Offline Ad Slot Scheduling} problem~\cite{FMNP} is defined as follows\index{Ad slot scheduling}\index{Offline ad slot scheduling}.
We have $n > 1$ bidders interested in clicks.  Each bidder $i$ has a
budget $B_i$ and a maximum cost-per-click (max-cpc) $m_i$.  Given a
number of clicks $c_i$, and a price per click $p$, the utility $u_i$
of bidder $i$ is $c_i$ if both the true max-cpc and the true budget
are satisfied, and $-\infty$ otherwise.  In other words, $u_i = c_i$
if $p \leq m_i$ and $c_i p \leq B_i$; and $u_i = -\infty$ otherwise.
We have $n'$ advertising slots where slot $i$ receives $D_i$ clicks
during the time interval $[0,1]$.  We assume $\DD_1 > \dots >
\DD_{n'}$.

In a {\em schedule}, each bidder is assigned to a set of (slot, time
interval) pairs $(j, [\sigma, \tau) )$, where $j \leq n'$ and $0 \leq
\sigma < \tau \leq 1$.  A {\em feasible schedule}\index{Feasible schedule in ad slot scheduling} is one where no
more than one bidder is assigned to a slot at any given time, and no
bidder is assigned to more than one slot at any given time.
(In other words, the intervals for a particular slot do not overlap, and
the intervals for a particular bidder do not overlap.)  
A feasible schedule can be applied as follows: when a user query comes
at some time $\sigma \in [0,1]$, the schedule for that time instant is
used to populate the ad slots.  If we assume that clicks come at a
constant rate throughout the interval $[0,1]$, the number of clicks a bidder
is expected to receive from a schedule is the sum of $(\tau - \sigma)
D_j$ over all pairs $(j, [\sigma, \tau) )$ in her
schedule.\footnote{All the results of~\cite{FMNP} generalize to the setting where
each bidder $i$ has a bidder-specific factor $\ctr_i$ in the click-through rate and thus receives $(\tau -
\sigma) \ctr_i D_j$ clicks (see Section~\ref{sec:conclusions}).  We
leave this out for clarity.}

A {\em mechanism} for {\em Offline Ad Slot Scheduling} takes as input
a declared budget $B_i$ and declared max-cpc (the ``bid'') $b_i$, and
returns a feasible schedule, as well as a price per click $p_i \leq
b_i$ for each bidder.  The schedule gives some number $c_i$ of clicks
to each bidder $i$ that must respect the budget at the given price;
i.e., we have $p_i c_i \leq B_i$.  
The {\em revenue} of a mechanism is $\sum_i p_i c_i$.  A
mechanism is {\em truthful} if it is a weakly dominant strategy to
declare one's true budget and max-cpc; i.e., for any bidder
$i$, given any set of bids and budgets declared by the other bidders,
declaring her true budget $B_i$ and max-cpc $m_i$ maximizes $u_i$.  In this setting, a (pure strategy) Nash equilibrium is a set of
declared bids and budgets such that no bidder wants to change her
declaration of bid or budget, given that all other declarations stay
fixed.  An {\em $\epsilon$-Nash equilibrium}\index{$\epsilon$-Nash equilibrium} is a set of bids and
budgets where no bidder can increase her $u_i$ by more than
$\epsilon$ by changing her bid or budget.

Throughout the presentation we assume some arbitrary
lexicographic ordering on the bidders, that does not necessarily match
the subscripts.  When we compare two bids $b_i$ and $b_{i'}$ we say
that $b_i \bidgt b_{i'}$ iff either $b_i > b_{i'}$, or $b_i = b_{i'}$
but $i$ occurs first lexicographically.

\smallskip

We comment that for this problem one is tempted to apply a {\em Fisher Market} model:
here $m$ divisible goods are available to $n$ buyers with money $B_i$,
and $u_{ij}(x)$ denotes $i$'s utility of receiving $x$ amount of good
$j$.  It is known~\cite{AD,EG,DPS} that under certain conditions a
vector of prices for goods exists (and can be found
efficiently~\cite{DPSV}) such that the {\em market clears}, in that
there is no surplus of goods, and all the money is spent.  The natural
way to apply a Fisher model to a slot auction is to regard the slots
as commodities and have the utilities be in proportion to the number
of clicks.  However this becomes problematic because there does not
seem to be a way to encode the scheduling constraints in the Fisher
model; this constraint could make an apparently ``market-clearing''\index{Market-clearing equilibrium}
equilibrium infeasible.

\subsection{Special Case: One Slot}
\label{sec:oneslot}

In this section we consider the case $k=1$, where there is only one
advertising slot, with some number $\DD := \DD_1$ of clicks.  A
truthful mechanism for this case is derived by first considering the two extreme
cases of infinite bids and infinite budgets.

Suppose all budgets $B_i = \infty$.  Then, our input amounts to bids
$b_1 \bidgt b_2 \bidgt \dots \bidgt b_n$.  The obvious mechanism is
simply to give all the clicks to the highest bidder.  We charge bidder
1 her full price $p_1 = b_1$.  A simple argument shows that reporting
the truth is a weakly dominant strategy for this mechanism.  Clearly
all bidders will report $b_i \leq \m_i$, since the price is set to
$b_i$ if they win.  The losing bidders cannot gain from decreasing
$b_i$.  The winning bidder can lower her price by lowering $b_i$, but
this will not gain her any more clicks, since she is already getting
all $D$ of them.

Now suppose all bids $b_i = \infty$; our input is just a set of
budgets $B_1, \dots, B_n$, and we need to allocate $D$ clicks, with no
ceiling on the per-click price.  Here we apply a simple rule known as
{\em proportional sharing}\index{Proportional sharing} (see~\cite{Kelly,JT}\footnote{Nguyen and
Tardos~\cite{NT} give a generalization of~\cite{JT} to general
polyhedral constraints, and also discuss the application to sponsored
search.  Both their bidding language and utility function differ
from~\cite{FMNP}, but it would be interesting to see if there are any
connections between their approach and~\cite{FMNP}.}): Let ${\cal B} =
\sum_i B_i$.  Now to each bidder $i$, allocate $({B_i}/{{\cal B}}) D$
clicks.  Set all prices the same: $p_i = p = {\cal B}/D$.  The
mechanism guarantees that each bidder exactly spends her budget, thus
no bidder will report $B'_i > B_i$.  Now suppose some bidder reports
$B_i' = B_i - \Delta$, for $\Delta >0$.  Then this bidder is allocated
$D({B_i - \Delta})/({{\cal B} - \Delta})$ clicks, which is less than
$D ({B_i}/{{\cal B}})$, since $n>1$ and all $B_i > 0$.

\heading{Greedy First-Price Mechanism.}
A natural mechanism for the general single-slot case is to solve the associated
``fractional knapsack''\index{Fractional knapsack problem} problem, and charge bidders their bid; i.e.,
starting with the highest bidder, greedily add bidders to the
allocation, charging them their bid, until all the clicks are
allocated.  We refer to this as the {\em greedy first-price}\index{GFP (Greedy First-Price) mechanism}\index{Greedy first-price (GFP) mechanism}
(GFP) mechanism. Though natural (and revenue-maximizing as a function of bids) this is
easily seen to be not truthful:

\begin{example}
\footnotesize
\cite{FMNP}
\label{ex:greedy-untruthful}
Suppose there are two bidders and $D = 120$ clicks.  
Bidder 1 has ($m_1 = \$2$, $B_1 = \$100$) and bidder 2 has ($m_2 = \$1$, $B_2 = \$50$).  
In the GFP mechanism, if both bidders tell the truth, then bidder 1
 gets 50 clicks for $\$2$ each, and 50 of the remaining 70 clicks 
go to bidder 2 for $\$1$ each.
However, if bidder 1 instead declares $b_1 = \$1 + \epsilon$, then she
gets (roughly) 100 clicks, and bidder 2 is left with (roughly) 20
clicks.
\end{example}

The problem here is that the high bidders can get away with bidding
lower, thus getting a lower price.  The difference between this and
the unlimited-budget case above is that a lower price now results in
more clicks.  It turns out that in equilibrium, this mechanism will
result in an allocation where a prefix of the top bidders are
allocated, but their prices equalize to (roughly) the lowest bid in
the prefix (as in the example above).  

\heading{The Price-Setting Mechanism.}
An equilibrium allocation of GFP can be computed directly via the
following mechanism, which~\cite{FMNP} refers to as the {\em price-setting (PS)
mechanism}.\index{Price-setting (PS) mechanism}\index{PS (Price-Setting) mechanism}  Essentially this is a descending price mechanism: the
price stops descending when the bidders willing to pay at that price
have enough budget to purchase all the clicks.  We have to be careful
at the moment a bidder is added to the pool of the willing bidders; if
this new bidder has a large enough budget, then suddenly the willing
bidders have {\em more} than enough budget to pay for all of the
clicks.  To compensate, the mechanism decreases this ``threshold''
bidder's effective budget until the clicks are paid for exactly.

\begin{mechanism}{Price-Setting (PS) Mechanism (Single Slot)~\cite{FMNP}}
$\bullet$ Assume wlog that $b_1 \bidgt b_2 \bidgt \dots \bidgt b_n \geq 0$.   \\
$\bullet$ Let $k$ be the first bidder such that $b_{k+1} \leq \sum_{i=1}^k B_i / D$.  Compute price $p = \min \{ \sum_{i=1}^k B_i / D, b_k \}$.\\
$\bullet$ Allocate $B_i / p$ clicks to each $i \leq k-1$.
Allocate $\hat{B}_k / p$ clicks to bidder $k$,
where $\hat{B}_k = p D - \sum_{i=1}^{k-1} B_i$.\\ 
\end{mechanism}

\begin{example}
\footnotesize
\cite{FMNP}
\label{ex:psm1}
Suppose there are three bidders with $b_1 = \$2$, $b_2 = \$1$, $b_3 =
\$0.25$ and $B_1 = \$100$, $B_2 = \$50$, $B_3 = \$80$, and $D = 300$
clicks.  
Running the PS mechanism, we get $k = 2$ since 
$B_1/D = 1/3 < b_2 = \$1$, but
$(B_1 + B_2)
/ D = \$0.50 \geq b_3 = \$0.25$.  The price is set to $\min \{ \$0.50,
\$1 \} = \$0.50$, and bidders 1 and 2 get 200 and 100 clicks at that
price, respectively.  There is no threshold bidder.
\end{example}

\begin{example}
\footnotesize
\cite{FMNP}
\label{ex:psm2}
Suppose now bidder 2 changes her bid to $b_2 = \$0.40$ (everything else remains the same as Example~\ref{ex:psm1}).   
We still get $k = 2$ since $B_1/D = 1/3 < b_2 = \$0.40$.  But now the
price is set to $\min \{ \$0.50, \$0.40 \} = \$0.40$, and bidders 1
and 2 get 250 and 50 clicks at that price, respectively.  Note that
bidder 2 is now a threshold bidder, does not use her entire budget, and gets fewer clicks.
\end{example}


\begin{theorem}
\label{thm:truth_single}
\rm{\cite{FMNP}}
The price-setting mechanism (single slot) is truthful.
\end{theorem}

\heading{Price-Setting Mechanism Computes Nash Equilibrium of GFP.}
Consider the greedy first-price auction in which the highest bidder
receives ${B_1}/{b_1}$ clicks, the second ${B_2}/{b_2}$ clicks and so
on, until the supply of $D$ clicks is exhausted.  It is immediate that
truthfully reporting budgets is a dominant strategy in this mechanism,
since when a bidder is considered, her reported budget is exhausted as
much as possible, at a fixed price.  However, reporting $b_i = m_i$ is
{\em not} a dominant strategy.  Nevertheless, it turns out that GFP
has an equilibrium whose outcome is (roughly) the same as the PS
mechanism.  One cannot show that there is a plain Nash equilibrium
because of the way ties are resolved lexicographically, so~\cite{FMNP}
proves instead that the bidders reach an $\epsilon$-Nash equilibrium:

\begin{theorem}
\label{thm:nash_single}
Suppose the PS mechanism is run on the truthful input,
resulting in price $p$ and clicks $c_1, \dots, c_n$ for each bidder.
Then, for any $\epsilon>0$ there is a pure-strategy $\epsilon$-Nash
equilibrium of the GFP mechanism where each bidder receives $c_i \pm
\epsilon$ clicks.
\end{theorem}

\subsection{Multiple Slots}
\label{sec:multislot}

Generalizing to multiple slots makes the scheduling constraint
nontrivial.  Now instead of splitting a pool of $D$ clicks
arbitrarily, we need to assign clicks that correspond to a feasible
schedule of bidders to slots.  The conditions under which this is
possible add a complexity that needs to be incorporated into
the mechanism.

As in the single-slot case it will be instructive to consider first
the cases of infinite bids or budgets.  Suppose all $B_i = \infty$.
In this case, the input consists of bids only $b_1 \bidgt b_2 \bidgt
\dots \bidgt b_n$.  Naturally, what we do here is rank by bid, and
allocate the slots to the bidders in that order.  Since each
budget is infinite, we can always set the prices $p_i$ equal to the
bids $b_i$. By the same logic as in the single-slot case, this is
easily seen to be truthful. In the other case, when $b_i = \infty$,
there is a lot more work to do.

Without loss of generality, we may assume the number of slots equals
the number of bids (i.e., $n' = n$); if this is not the case, then we
add dummy bidders with $B_i = b_i = 0$, or dummy slots with $D_i = 0$,
as appropriate.  We keep this assumption for the remainder of the
section.

\heading{Assigning Slots Using a Classical Scheduling Algorithm.}
  First we give an important lemma that characterizes the
conditions under which a set of bidders can be allocated to a set of
slots, which turns out to be just a restatement of a classical
result~\cite{HLS} from scheduling theory.

\begin{lemma}\label{lemma:condition}
\rm{\cite{HLS,FMNP}}
Suppose we would like to assign an arbitrary set $\{1, \dots, k\}$ of
bidders to a set of slots $\{1, \dots, k\}$ with $D_1 > \dots >
D_k$.  Then, a click allocation $c_1 \geq ...\geq c_k$ is feasible iff
\begin{eqnarray}\label{eq:condition}
c_1 + \dots + c_\ell \leq D_1 + \dots + D_\ell \quad\textrm{ for all } \ell=1,...,k.
\end{eqnarray}
\end{lemma}
\begin{proof}
In scheduling theory, we say a {\em job} with {\em service
requirement} $x$ is a task that needs $x / s$ units of time to
complete on a {\em machine} with {\em speed} $s$.
The question of whether there is a feasible allocation is equivalent
to the following scheduling problem: Given $k$ jobs with service
requirements $x_i = c_i$, and $k$ machines with
speeds $s_i = D_i$, is there a schedule of jobs to
machines (with preemption allowed) that completes in one unit of time?

As shown in~\cite{HLS,GS}, the optimal schedule for this problem
(a.k.a. $Q \: | \: \textit{pmtn} \: | \: C_{\max}$) can be found
efficiently by the {\em level algorithm},\index{Level algorithm}
and the schedule completes in time $\max_{\ell \leq k}
\{{\sum_{i=1}^\ell x_i}/{\sum_{i=1}^\ell s_i}\}$.  Thus, the
conditions of the lemma are exactly the conditions under which the
schedule completes in one unit of time.\qed
\end{proof}

\heading{A Multiple-Slot Budgets-Only Mechanism.}
The mechanism in~\cite{FMNP} is roughly a
descending-price mechanism\index{Descending-price mechanism} where we decrease the price until a prefix
of budgets fits tightly into a prefix of positions at that price, whereupon we
allocate that prefix, and continue to decrease the price for the
remaining bidders.  More formally, it can be written as follows:

\begin{mechanism}{Price-Setting Mechanism (Multiple Slots, Budgets Only)~\cite{FMNP}}
$\bullet$ If all $D_i = 0$, assign bidders to slots arbitrarily and exit.\\
$\bullet$ Sort the bidders by budget and assume wlog that $B_1 \geq B_2 \geq ... \geq B_n$. \\
$\bullet$ Define $r_\ell = {\sum_{i=1}^{\ell} B_i}/{\sum_{i=1}^\ell D_i}$. 
Set price $p = \max_\ell r_\ell$. \\
$\bullet$
Let $\ell^*$ be the largest $\ell$ such that $r_\ell = p$.  Allocate
slots $\{1, \dots \ell^*\}$ to bidders $\{1, \dots, \ell^*\}$ at price
$p$, using all of their budgets; i.e., $c_i = B_i / p$. \\ 
$\bullet$ Repeat the steps above on the remaining bidders and slots until all slots are allocated. \\ 
\end{mechanism}

\begin{figure}
\psfrag{#bidder#}{\bf Bidder} 
\psfrag{#budget#}{\bf Budget}

\psfrag{#b1#}{1}
\psfrag{#b2#}{2}
\psfrag{#b3#}{3}
\psfrag{#b4#}{4}

\psfrag{#B1#}{$\$80$}
\psfrag{#B2#}{$\$70$}
\psfrag{#B3#}{$\$20$}
\psfrag{#B4#}{$\$1$}

\psfrag{#35#}{$3/5$}
\psfrag{#25#}{$2/5$}
\psfrag{#2021#}{$20/21$}
\psfrag{#121#}{$1/21$}

\psfrag{#line1#}{$D_1 = 100$}
\psfrag{#line2#}{$D_2 = 50$}
\psfrag{#line3#}{$D_3 = 25$}
\psfrag{#line4#}{$D_4 = 0$}

\psfrag{#block1#}{$p_1 = \$1.00$}
\psfrag{#block2#}{$p_2 = \$0.84$}
\includegraphics[width=4.5in]{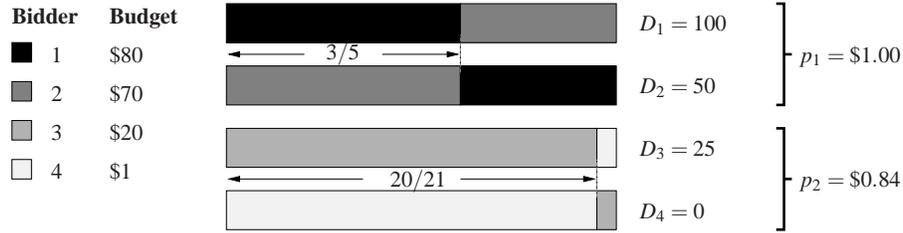}
\caption{
\footnotesize
An example of the PS mechanism (multiple slots,
budgets only). 
The first application of Find-Price-Block computes
$r_1 = B_1 / D_1 = 80/100$, 
$r_2 = (B_1 + B_2) / (D_1 + D_2) = 150 / 150$, 
$r_3 = (B_1 + B_2 + B_3) /  (D_1 + D_2 + D_3) = 170 / 175$,
$r_4 = (B_1 + B_2 + B_3 + B_4) /  (D_1 + D_2 + D_3 + D_4) = 171 / 175$.
Since $r_2$ is largest, the top two slots make up the first price
block with a price $p_1 = r_2 = \$1$; bidder 1 gets $80$ clicks and
bidder 2 gets $70$ clicks, using the schedule as shown.
In the second price block, we get $B_3 / D_3 = 20/25$ and $(B_3 +
B_4) / (D_3 + D_4) = 21/25$.  Thus $p_2$ is set to $21/25 =
\$0.84$, bidder $3$ gets $500 / 21$ clicks and bidder $4$ gets
$25/21$ clicks, using the schedule as shown.
}
\label{ex:1}
\vspace{-.2in}
\end{figure}

\noindent Note that the allocation step is always possible since for
all $\ell \leq \ell^*$, we have $p \geq r_\ell = {\sum_{i=1}^{\ell}
B_i}/{\sum_{i=1}^\ell D_i}$, which rewritten is $\sum_{i=1}^\ell c_i
\leq \sum_{i=1}^\ell D_i$, and so we can apply
Lemma~\ref{lemma:condition}.  An example run of the price-setting
mechanism is shown in Figure~\ref{ex:1}.

\begin{theorem}\label{thm:monotonicity}
\rm{\cite{FMNP}}
The price-setting mechanism (multi-slot, budgets only) is truthful. 
\end{theorem}



\heading{The Price-Setting Mechanism (General Case).}  The
generalization of the multiple-slot PS mechanism to use both bids and budgets combines the ideas from the
bids-and-budgets version of the single slot mechanism with the
budgets-only version of the multiple-slot mechanism.  As our price
descends, we maintain a set of ``active'' bidders with bids at or
above this price, as in the single-slot mechanism.  These active
bidders are kept ranked by {\em budget}, and when the price reaches
the point where a prefix of bidders fits into a prefix of slots (as in
the budgets-only mechanism) we allocate them and repeat.  As in the
single-slot case, we must be careful when a bidder enters the active
set and suddenly causes an over-fit; in this case we again reduce the
budget of this ``threshold'' bidder until it fits.  For details on
this mechanism and a proof that it is also truthful, we refer the
reader to the paper~\cite{FMNP}.

\heading{Greedy First-Price Mechanism for Multiple Slots.}
In the multiple-slot case, as in the single-slot case, there is a natural
{\em greedy first-price} (GFP)\index{Greedy first-price (GFP) mechanism}\index{GFP (Greedy first-price) mechanism} mechanism when the bidding language includes
both bids and budgets: Order the bidders by bid $b_1 \bidgt b_2 \bidgt \dots \bidgt
b_n$. Starting from the highest bidder, for each bidder $i$ compute
the maximum possible number of clicks $c_i$ that one could allocate to
bidder $i$ at price $b_i$, given the budget constraint $B_i$ and the
commitments to previous bidders $c_1, \dots, c_{i-1}$.  This reduces
to the ``fractional knapsack'' problem in the single-slot case, and so
one would hope that it maximizes revenue for the given bids and
budgets, as in the single-slot case.  This is not immediately clear,
but does turn out to be true (see~\cite{FMNP} for details).
As in the single-slot case, the GFP mechanism is not a truthful
mechanism.  However,~\cite{FMNP} give a generalization of Theorem~\ref{thm:nash_single}
showing that
the multiple-slot GFP mechanism does have a pure-strategy
equilibrium, and that equilibrium has prices and allocation equivalent
to the multiple-slot price setting mechanism.

\ignore{
\heading{Greedy is Revenue-Maximizing.} 
Consider a revenue-maximizing schedule that respects both bids and
budgets.  In this allocation, we can assume wlog that each bidder $i$
is charged exactly $b_i$ per click, since otherwise the allocation can
increase the price for bidder $i$, reduce $c_i$ and remain feasible.
Thus, by Lemma~\ref{lemma:condition}, we can find a revenue-maximizing
schedule $\carr^* = (c^*_1, \dots, c^*_n)$ by maximizing
$\sum_{i}{b_i c_i}$ subject to $c_i \leq B_i / b_i$
and
$
c_1 + \dots + c_\ell \leq D_1 + \dots + D_\ell$ for all $\ell=1,...,n.
$

\begin{theorem}\label{thm:equivalence}
The GFP auction gives a revenue-maximizing schedule.
\end{theorem}

\heading{Price-Setting Mechanism is a Nash Equilibrium of the Greedy First Price Mechanism.}
We note that truthfully reporting one's budget is a weakly dominant
strategy in GFP, since when a bidder is considered for allocation,
their budget is exhausted at a fixed price, subject to a cap on the
number of clicks they can get.  Reporting one's bid truthfully is not a
dominant strategy, but we can still show that there is an
$\epsilon$-Nash equilibrium whose outcome is arbitrarily close to the
PS mechanism.

\begin{theorem}
\label{thm:nash}
Suppose the PS mechanism is run on the truthful input,
resulting in clicks $c_1, \dots, c_n$ for each bidder.  Then, for any
$\epsilon>0$ there is a pure-strategy $\epsilon$-Nash equilibrium of the GFP
mechanism where each bidder receives $c_i \pm \epsilon$ clicks.
\end{theorem}

}

\subsection{Extensions}
\label{sec:conclusions}

There are several natural generalizations of the 
{\em Online Ad Slot Scheduling} problem where it would be interesting
to extend or apply the results of~\cite{FMNP}:
\begin{itemize}
\item
{\it Click-through rates.} 
To incorporate ad-specific click-through rates $\ctr_i$ into this model,
we would say that a bidder $i$ assigned to slot $j$ for a time period of
length $\tau - \sigma$ would receive $(\tau - \sigma) \ctr_i D_j$ clicks.  All the
results of~\cite{FMNP} can be generalized to this setting by simply scaling the bids
using $b'_i = b_i \ctr_i$.  However, now the mechanism does
not necessarily prefer more {\em efficient} solutions; i.e., ones that generate
more overall clicks.  It would be interesting to analyze a possible
tradeoff between efficiency and revenue in this setting.  

\item
{\it Multiple Keywords.} To model multiple keywords in this model, we
could say that each query $q$ had its own set of click totals $D_{q,
1} \dots D_{q, n}$, and each bidder is interested in a subset of
queries.  The greedy first-price mechanism is easily generalized to
this case: maximally allocate clicks to bidders in order of their bid
$b_i$ (at price $b_i$) while respecting the budgets, the query
preferences, and the click commitments to previous bidders.  It would
not be surprising if there was an equilibrium of this extension of 
the greedy mechanism that could
be computed directly with a generalization of the PS
mechanism.

\item
{\it Online queries, uncertain supply.} In sponsored search,
allocations must be made online in response to user queries, and some
of the previous literature has focused on this aspect of the problem
(e.g., \cite{MSVV,MNS}). Perhaps the ideas from~\cite{FMNP} could be
used to help make online allocation decisions using (unreliable)
estimates of the supply, a setting considered in~\cite{MNS}, with game-theoretic
considerations.
\end{itemize}

\section{The User's Point of View: a Markov Model for Clicks}
\label{sec:markov}

In the GSP auction, by fixing the sort order, we leave out an important
third party in sponsored search; i.e., the {\em search engine user}.
Unfortunately, there is very little guidance on this in the
literature, even though the user's behavior is the essential
ingredient that defines
{\em the commodity} the advertisers are bidding on, and its value.  In~\cite{AFMP08}
 a different framework is suggested for principled understanding of
sponsored search auctions:
\begin{itemize}
\item Define a suitable probabilistic model for search engine user
behavior upon being presented the ads.
\item Once this model is fixed, ask the traditional mechanism design
questions of how do assign the ads to slots, and how to price them.
\item Analyze the given mechanism from the perspective of the bidders
(e.g., strategies) and the search engine (e.g., user satisfaction,
efficiency and revenue).
\end{itemize}

There are certain well-accepted observations about the user's
interaction with the sponsored search ads that should inform the
model: 
\begin{itemize}
\item
 The higher the ad is on the page, the more clicks it
gets. 
\item
The ``better'' the ad is, the more clicks it gets,
where the ``goodness'' of an ad is related to the inherent
quality of the ad, and how well it matches the user's query.
\end{itemize}
These properties govern not only how the auction is run but also how
advertisers think about their bidding strategy (they prefer to appear
higher and get more clicks).  Thus it is important for an auction to
have what we call {\em intuitive bidding}: a higher bid translates to
a higher position and more clicks.


In~\cite{AFMP08}, a natural Markov model is proposed for user clicks,
taking the above observations into account.   An algorithm is given to
determine an optimal assignment of ads to positions in terms of
economic efficiency. Together with VCG pricing, this gives a truthful
auction.  They further show that the optimal assignment under
this model has certain monotonicity properties that allow for
intuitive bidding. In what follows, we will describe these
contributions in more detail.

\heading{Modeling the Search Engine User.}
Previous work on sponsored search has (implicitly) modeled the user
using two types of parameters: ad-specific click-through rates
$\userctr_i$ and position-specific visibility factors
$\pn_j$.
There are some intuitive user behavior models that express overall
click-through probabilities in terms of these parameters.
One possibility is ``for each position $j$ {\em independently}, the
user looks at the ad $i$ in that position with probability $\pn_j$
then clicks on the ad with probability $\userctr_i$.''  Alternatively:
``The user picks a {\em single} position according to the distribution
implied by the $\pn_j$'s, and then clicks on the ad $i$ in that
position with probability $\userctr_i$.''  Under both these models, it
follows that the probability of an ad $i$ in position $j$ receiving a
click is equal to $\userctr_i \pn_j$, which is the so-called {\em
separability} assumption (see~\cite{AGM} or the discussion in
Section~\ref{sec:existing}).  From separability it follows that GSP
ordering of ads will be suitable, because GSP ordering maximizes the
total advertiser value on the page.

In both these models there is no reason {\em a priori}
that the position factors $\pn_j$ should be decreasing; this is simply
imposed because it makes sense, and it is verifiable empirically. 
Also, both suggested models assume that the probability of an ad getting
clicked is independent of {\em other ads} that appear with it on the
page, an assumption made without much justification.  It is hard to
imagine that seeing an ad, perhaps followed by a click, has no effect
on the subsequent behavior of the user.  

In designing a user model, we would like to have the monotonicity of
the positions arise naturally.  Also, each ad should have parameters
dictating their effect on the user both in terms of clicking on that
ad, as well as looking at other ads.  In~\cite{AFMP08}, a model is proposed of a
user who starts to scan the list of ads from the top, and makes
decisions (about whether to click, continue scanning, or give up
altogether) based on what he sees.  More specifically, the
user is modeled as the following Markov process: ``Begin scanning the ads from
the top down.  When position $j$ is reached, click on the ad $i$ with
probability $\userctr_i$.  Continue scanning with probability $q_i$.''  In
this model, if we try to write the click probability of an ad $i$ in
position $j$ as $\userctr_i \pn_j$, we get that $\pn_j = \Pi_{i' \in A}
q_{i'}$, where $A$ is the set of ads placed above\footnote{Throughout the
section, we will often refer to a position or an ad being ``higher'' or
``above'' another position or ad; this means that it is earlier on the
list, and is looked at first by the user.}
 position $j$.  Thus
the ``position factor'' in the click probability decreases with
position, and does so naturally from the model.  Also note that we do
not have separability anymore, since $\pn_j$ depends on which ads
are above position $j$. Consequently, it can be shown that GSP
assignment of ads is no longer the most efficient.

\heading{Auction with Markovian users.}  Given this new user model, we
can now ask what the best assignment is of ads to slots.  In~\cite{AFMP08}, 
the most efficient assignment is studied; i.e., the one that maximizes
total advertiser value derived from user clicks.  It turns out that
the structure of this assignment is different than that of GSP, and
indeed is more sophisticated than any simple ranking.  The presence of
the $q_i$'s requires a delicate tradeoff between the click probability
of an ad and its effect on the slots below it.  In~\cite{AFMP08},
 certain structural properties of the optimal assignment are identified and
used to find such an optimal assignment efficiently, not only in
polynomial time, but in near-linear time.  Given this algorithm, a
natural candidate for pricing is VCG~\cite{V,C,G}, which is clearly
truthful in this setting, at least under a profit-maximizing utility.

\heading{Intuitive Bidding.}\index{Intuitive bidding in sponsored search}
One of the reasons why GSP is successful is perhaps because bidding
strategy is intuitive: Under GSP ranking, if an advertiser bids more,
they get to a higher position, and consequently, if they bid more,
their click probability increases.  Now that we have defined a more
sophisticated assignment function, even though VCG pricing is truthful,
the auction still may not have these intuitive properties.  The main technical
result in~\cite{AFMP08} is to show that in the Markov user  model, if a mechanism uses the most
efficient assignment, indeed position and click probabilities are
monotonic in an ad's bid (with all other bids fixed), thus preserving
this important property.  While not surprising, position-monotonicity
turns out to be rather involved to prove, requiring some delicate
combinatorial arguments, and insights into the optimal substructure of
bidder assignments.

\medskip
In summary, sponsored search auctions are a three party
process which can be studied by modeling the behavior of users first and then designing
suitable mechanisms to affect the game theory between the advertiser
and the search engine.  The work of~\cite{AFMP08} sheds some light on the intricate
connection between the user models and the mechanisms; for example,
the sort order of GSP that is currently popular (sort by
$b_i\userctr_i$) is not optimal under the Markov user model.

\subsection{A Simple Markov User Click Model}
\label{sec:markovmodel}
We consider a sponsored search auction with $n$ bidders $\bidders =
\{1, \dots, n\}$ and $k$ positions.  We will also refer to ``ad $i$,''
meaning the advertisement submitted by bidder $i$.  Each bidder $i \in
\bidders$ has two parameters, $\userctr_i$ and $q_i$.  The
click-through-rate $\userctr_i$ is the probability that a user will click
on ad $i$, given that they {\em look} at it.  The continuation
probability $q_i$ is the probability that a user will look at the next
ad in a list, given that they look at ad $i$\index{Markovian user model}.  

Each bidder submits a bid $b_i$ to the auction, representing the
amount that they value a click.  The quantity $\userctr_i b_i$ then
represents the value of an ``impression,'' i.e., how much they value a
user looking at their ad.  This is commonly referred to as their
``ecpm.''\footnote{The acronym ecpm stands for ``expected cost per
thousand'' impressions, where M is the roman numeral for one
thousand. We will drop the factor of one thousand and refer to $\userctr_i
b_i$ as the ``ecpm.''}\index{ecpm (Expected Cost Per Thousand [Impressions])}\index{Expected cost per thousand [impressions] (ecpm)} Throughout, we will use the notation $\eff_i = \userctr_i
b_i$ for convenience.

Given an assignment $(x_1, \dots, x_k)$ of bidders to the $k$
positions, the user looks at the first ad $x_1$, clicks on it with
probability $\userctr_{x_1}$, and then continues looking with probability
$q_{x_1}$.\footnote{The click event and the continuation event could
in principle have some correlation, and the results mentioned here will still hold.}
  This is repeated with the second bidder, etc., until the
last ad is reached, or some continuation test has failed.  Thus the
overall expected value of the assignment to the bidders is
$$
\eff_{x_1} + q_{x_1} (\eff_{x_2} + q_{x_2}(\eff_{x_3} + q_{x_3}(\dots q_{x_{n'-1}}(\eff_{x_n})))).
$$

Now that we have defined the user model, and characterized the value
of an assignment in that model, we can now define a new auction
mechanism:  First, the search engine computes an assignment of ads to
positions that maximizes the overall expected value.  Given this
assignment, prices can then be computed using VCG~\cite{V,C,G}; for each
assigned bidder we compute the change in others' value if that bidder
were to disappear.  This assures truthful reporting of bids under a
profit-maximizing utility function.

\subsection{Properties of Optimal Assignments for Markovian Users}
\label{sec:props}

Since the optimal assignment used by the mechanism is no longer simple
ranking by ecpm, it is essential to understand the structure of this
assignment.  This understanding will allow us to compute the
assignment more efficiently, and prove some important game-theoretic
properties of the mechanism.

It turns out that the quantity $\eff_i / (1-q_i)$, which we will refer to
as the ``adjusted ecpm (a-ecpm),''\index{Adjusted ecpm (a-ecpm)}\index{a-ecpm (Adjusted ecpm)} plays a central role in this model.
Intuitively, this quantity is the impression value adjusted by the
negative effect this ad has on the ads below it.  We use $\aecpm_i =
\eff_i / (1-q_i)$ for convenience.  The following theorem tells us how to
assign a set of $k$ selected ads to the $k$ positions:

\begin{theorem}
\rm{\cite{AFMP08}}
In the most efficient assignment, the ads that are placed are sorted in decreasing order of
adjusted ecpm 
$
\aecpm_i = \eff_i / (1 - q_i)
$.
\label{thm:rank}
\end{theorem}

\newcommand{\effnum}{{\hat{\eff}}}

While this theorem tells us how to sort the ads selected, it does not
tell us {\em which} $k$ ads to select.  One is tempted to say that
choosing the top $k$ ads by a-ecpm would do the trick; however
the following example proves otherwise:

\begin{example}
\footnotesize
\cite{AFMP08}
\label{mex:1}
Suppose we have three bidders and two slots, and the bidders have the following parameters:

\smallskip
\begin{center}
\begin{tabular}{rrrr}
Bidder & $\eff_i$  & $q_i$ & $\aecpm_i = \eff_i / (1-q_i)$ \\ \hline
1      & \$1    & .75   & 4      \\ 
2      & \$2    & .2    & 2.5    \\ 
3      & \$0.85 & .8    & 4.25   \\ 
\end{tabular}
\end{center}
\smallskip

Let's consider some possible assignments and their efficiency.  If we
use simple ranking by ecpm $\eff_i$, we get the assignment $(2,1)$, which has
efficiency $\$2 + .2 (\$1) = \$2.20$.  If we use simple ranking by a-ecpm $a_i$ we
get the assignment $(3,1)$ with efficiency $\$0.85 + .8 (\$1) =
\$1.65$.  It turns out that the optimal assignment is $(1,2)$ with
efficiency $\$1 + .75 (\$2) = \$2.50$.  The assigned bidders are
ordered by a-ecpm in the assignment, but are not the top 2 bidders by
a-ecpm.

Now suppose we have the same set of bidders, but now we have three
slots.  The optimal assignment in this case is $(3,1,2)$; note how
bidder 3 goes from being unassigned to being assigned the first
position.
\end{example}


In classical sponsored search with simple ranking, a bidder $j$ can
dominate another bidder $i$ by having higher ecpm; i.e., bidder $j$
will always appear whenever $i$ does, and in a higher position.
Example~\ref{mex:1} above shows that having a higher ecpm (or a-ecpm) does not allow a bidder to dominate another bidder in our new
model.  However, we show that if she has higher ecpm
{\em and} a-ecpm, then this does suffice.  This is not only
interesting in its own right, it is essential for proving deeper
structural properties.

\begin{theorem} 
\rm{\cite{AFMP08}}
\label{lemma:sub}
For all bidders $i$ in an optimal assignment, if some bidder $j$ is not in the
assignment, and $\aecpm_j \geq \aecpm_i$ and $\eff_j
\geq \eff_i$, then we may substitute $j$ for $i$, and the assignment is
no worse.
\end{theorem}

The following 
 theorem shows some
subset structure between optimal assignments to different numbers of slots.
This theorem is used to prove position monotonicity, and is an essential ingredient of
the more efficient algorithm for finding the optimal assignment.  Let $\opt(C,j)$ denote the set of all optimal
solutions for filling $j$ positions with bidders from the set $C$.

\begin{theorem} 
\label{thm:subsets}
\rm{\cite{AFMP08}}
Let $j \in \{1, \dots, k\}$ be some number of positions, and let $C$
be an arbitrary set of bidders.  Then, for all $S \in \opt(C,j-1)$,
there is some $S' \in \opt(C,j)$ where $S' \supset S$.
\end{theorem}


Finally, we state a main technical theorem of~\cite{AFMP08}, which shows that bidding
is intuitive under a mechanism that maximizes value in the Markovian
model.

\begin{theorem}
\label{thm:monotonic}
\rm{\cite{AFMP08}}
With all other bids fixed, the probability of receiving a click in the
optimal solution is non-decreasing in one's bid.  In addition, the
position of a particular bidder in the optimal solution is monotonic
that bidder's bid.
\end{theorem}

This theorem, whose proof relies on all the previous results in this
section, implies that from the perspective of a bidder participating
in the auction, all the complexities of the underlying assignment
still do not interfere with the intuitive nature of bidding; if you
bid more, you still get more clicks, and get to a higher position.

\subsection{Computing the Optimal Assignment}

A simple algorithm for computing the optimal assignment proceeds as follows.  First, sort the ads in decreasing
order of a-ecpm in time $O(n \log n)$.  Then, let $F(i,j)$ be
the efficiency obtained (given that you reach slot $j$) by filling
slots $(j, \dots, k)$ with bidders from the set $\{i, \dots, n\}$.  We
get the following recurrence:
$$
F(i,j) = \max (F(i+1, j+1) q_i + \eff_i, F(i+1, j)).
$$ 
Solving this recurrence for $F(1,1)$ yields the optimal assignment,
and can be done in $O(nk)$ time. 
Using the properties about optimal assignments proved in the previous section, this can be improved  to

\begin{theorem}
\rm{\cite{AFMP08}}
Consider the auction with $n$ Markovian bidders and $k$ slots. 
There is an optimal assignment which can be determined in $O(n\log n + k^2 \log^2 n)$ time.
\end{theorem}

\section{Open Issues}

We emphasize three open directions, besides the various game-theoretic and algorithmic
open problems already proposed so far. 

\begin{itemize}
\item
{\em Estimating Parameters.}
In order to run the basic auction and its extensions, the search companies need to estimate 
a number of parameters: ctr, position-specific factors, minimum bidder-specific reserve prices, etc. In 
addition, for operational reasons, search engines have to provide traffic estimates to potential advertisers,
that is, for each keyword, they need to show landscape functions such as the ones in Section 3.
 An open research problem is, given a log of search and ad traffic over a significant 
period of time,  design and validate efficient learning methods for estimating these parameters, and 
perhaps, even identify the models that fit the variation of these parameters over time. This is a 
significant research challenge since these parameters have intricate dependencies, and 
in addition, there is a long tail effect in the logs, that is, there is a significant amount of rare queries and keywords, 
as well as rare clicks for particular keywords. 

\item
{\em Grand Simulation.}
In order for the academic world to develop intuition into the world of sponsored search auctions and the associated
dynamics, we need a grand simulation platform that can generate search traffic, ad inventories, ad clicks, and market 
specifics at the ``Internet scale'' that search engines face. Such a platform will help us understand the many 
tradeoffs: increasing keywords vs increasing budget for a campaign, making better bids vs choosing different search engines, 
choosing to bid for impressions vs clicks vs action, etc. Some auction programs are currently available\footnote{For example,
see \url{http://www.hss.caltech.edu/~jkg/jAuctions.html}}, but a systematic, large scale effort by academia will have
tremendous impact for research.

\item
{\em Grand Models.}
In general, we need more detailed models for the behavior of users, advertisers as well as the impact of the 
search engine design on them.  We described a highly preliminary effort here in which the users were Markovian, but
more powerful models will also be of great interest. For
example, a small
extension is to make the continuation probability $q_i$ a
function of location as well, which makes the optimization problem
more difficult.  One can also generalize the Markov model to handle
arbitrary configurations of ads on a web page (not necessarily the linear order in current
search results page), or to allow various other user states (such as
navigating a {\em landing} page,\index{Landing page in sponsored search} that is the page that is the target of an ad).  Finally, since page layout can be
performed dynamically, we could ask what would happen if the layout of
a web page were a part of the model, which would combine both users as well as the search engine into a model. 
In general, there may be grand, unified models that capture the relationship between all the three parties in sponsored search.
\end{itemize}

\section{Concluding Remarks}

We have discussed algorithmic and game-theoretic issues in auctions for sponsored search.

Auctions are used for other products in Internet advertising, for example 
Google's {\em AdSense},\index{Google}\index{AdSense (Google)} where an Internet publisher (like an online newspaper) can sign up with an ad network (in this case Google)
to place ads on their site.
 Here, an additional aspect of the problem from the auctioneer's perspective is how to {\em target}
ads, that is, how to choose the keywords from the surrounding context, to run auctions
like the ones we discussed thus far. This also introduces the fourth player in the game, i.e.,
the {\em publisher}, and consequently, the game theory is more intricate and largely unexplored.

Internet ads like sponsored search or AdSense may combine different types of ads, i.e., text, image or video ads. Each has its
own specifications in terms of dimensions, user engagement and effectiveness. How to combine them
into a unified auction is an interesting challenge. 

Beyond Internet ads, the Internet medium is also used for enabling ads
in traditional media including TV, Radio, Print etc. In such
cases, the auction problems may take on a richer combinatorial
component, and also, a component based on ability to reserve ad slots
ahead of time. The resulting algorithmic and game-theoretic problems
are largely unexplored in the research community.

\ignore{
  Of course, auctions
are also used on the Internet for things besides advertising, and more
directly for pricing, for example, on eBay. This note has focused on
sponsored search only.
}

\section{Acknowledgements}

We gratefully thank numerous engineers and researchers at Google.

\bibliographystyle{plain}
\bibliography{sponsored_search}

\end{document}